\documentclass[journal=jacsat,manuscript=article]{achemso}

\usepackage[version=3]{mhchem} 
\usepackage{epstopdf}
\usepackage{graphicx}

\author{N. J. Lambert}
\email{nl249@cam.ac.uk}
\author{M. Edwards}
\author{C. Ciccarelli}
\author{A. J. Ferguson}
\email{ajf1006@cam.ac.uk}
\affiliation{Cavendish Laboratory, J.\ J.\ Thomson Avenue, Cambridge, CB3 0HE, United Kingdom}

\title{A charge parity ammeter}


\begin{document}
\begin{abstract}
A metallic double-dot is measured with radio frequency reflectometry. Changes in the total electron number of the double-dot are determined via single electron tunnelling contributions to the complex electrical impedance. Electron counting experiments are performed by monitoring the impedance, demonstrating operation of a single electron ammeter without the need for external charge detection.\end{abstract}

Single electron ammeters count each individual electron that passes. They enable the detection of tiny ($I < \mathrm{aA}$) currents\cite{Bylander2004}; show the real-time dynamics of single electron tunnelling\cite{Lu2003,Guttinger2011}; and may be of use in metrology for the definition of the Ampere\cite{Keller2008,Giblin2010,Keller1996,Fricke2011,Yamahata2011}. These devices are made possible by high sensitivity, high bandwidth charge detection which can register the passage of an electron through multiple island single electron devices\cite{Fujisawa2006}. Typically the charge detector employed is the single electron transistor\cite{Fujisawa2004} or quantum point contact\cite{Schleser2004,Vandersypen2004,Gustavsson2007}, which may be operated in a radio-frequency impedance matching circuit to increase both bandwidth and sensitivity\cite{Schoelkopf1998a,Qin2006,Reilly2007}. In this Letter we show that use of external electrometry is unnecessary, simplifying the fabrication of a single electron ammeter. Rather, the internal tunnelling processes within a metallic double dot can be used to herald the passage of single electrons through the system, so the double dot acts as its own electrometer. Specifically, we demonstrate real-time measurements of the relative charge parity of the metallic double dot by probing the complex impedance of the device itself.

When any of the capacitors in a single electron device is driven by an alternating potential, periodic tunnelling of electrons can occur, resulting in an alternating current. For example, if an alternating potential is applied to the gate of a single electron box, a current is periodically driven across the tunnel barrier. If the drive frequency is sufficiently high that electron tunnelling does not happen adiabatically, energy is dissipated each half cycle leading to an effective resistance\cite{Persson2010} - the `Sisyphus resistance'. In general the driven tunnel current is not in phase with the applied alternating potential and both real and imaginary components of the Sisyphus effect need to be considered\cite{Ciccarelli2011,Ota2010,Ashoori1992a}. In this Letter we refer to the real and imaginary Sisyphus impedances as $Z_{S}^{Re}$ and $Z_{S}^{Im}$ respectively. They are measured here by their dissipative and dispersive effects on a radio-frequency resonant circuit. They can be observed in any single electron system; in this sense they are dissimilar to the `quantum capacitance' due to the change in bandstructure curvature close to an anti-crossing between two levels, which is present in, for example, Cooper pair boxes\cite{Duty2005} and double quantum dots\cite{Petersson2010,Schroer2012}.

\begin{figure}
\includegraphics{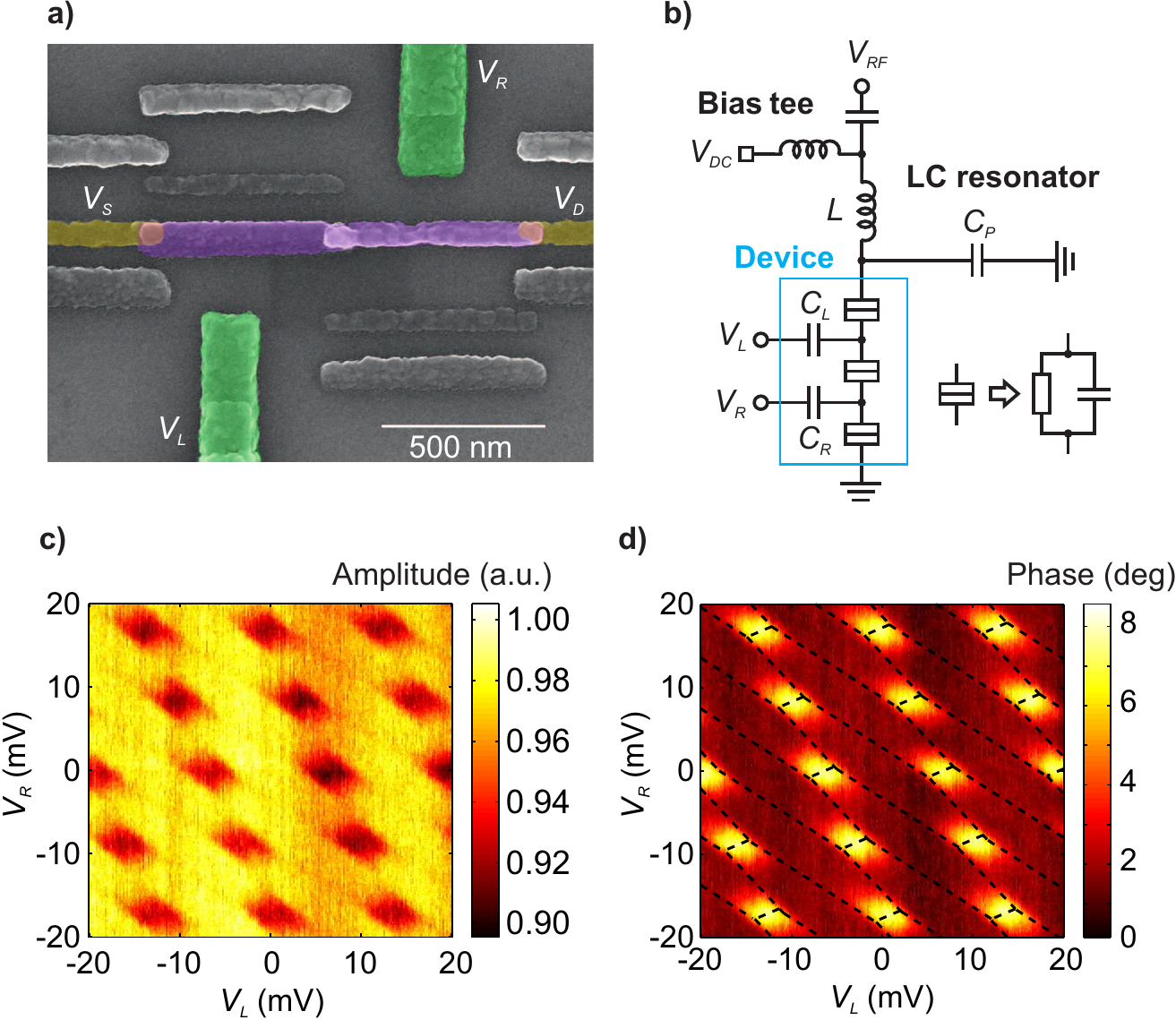}
\caption{(a) False colour SEM of the device, with source and drain leads (yellow; $V_{s}$ and $V_{d}$) and electrostatic gates (green; $V_{R}$, $V_{L}$) highlighted. Islands are shown in purple. (b) Experimental wiring, with the equivalent circuit for the device embedded in a resonant circuit. A bias tee allows the application of a d.c.\ source-drain voltage ($V_{ds}$) in addition to the r.f.\ drive. (c) The amplitude of the reflected r.f.\ signal for the LR device as a function of $V_{L}$ and $V_{R}$. For each column of (c) and (d), $V_{L}$ is fixed, while $V_{R}$ is swept and the average of 100 measurements taken.  (d) The phase of the reflected signal as a function of $V_{L}$ and $V_{R}$. Zero phase is chosen to be away from charge degeneracy points, and the boundaries of the charge stability diagram are highlighted.}
\label{fig1}
\end{figure}

Our metallic double dot (fig.\ 1a) is defined by multiple angle shadow mask evaporation\cite{Fulton1987}. It comprises two micro-scale aluminium islands in series, with each island capacitatively coupled to an electrostatic gate, allowing individual control of their chemical potentials. The two islands are separated from each other by a nano-scale alumina tunnel barrier created by controlled oxidation. From previous experiments we find this to give a barrier resistance of around $30$ k$\Omega$ for junctions of this size. Aluminium source and drain leads allow electrical contact to the islands. We fabricate samples with both low and high lead-dot tunnel junction resistances (here called LR and HR devices). The LR device has a total resistance of $\sim9.6$ M$\Omega$. Assuming symmetric tunnel barriers, the lead dot resistances are therefore $\sim4.8$ M$\Omega$. The HR device, however, has much higher resistances of $\sim54$ T$\Omega$.

All our measurements are made at milli-Kelvin temperature in a dilution refrigerator with a magnetic field (400 mT) applied to suppress the superconductivity of the aluminium. The device is embedded in a radio frequency resonant circuit (fig.\ 1b) which comprises a chip inductor ($L = 470$ nH) and a parasitic capacitance ($C_p \sim 0.4$ pF). The circuit is placed in a reflectometry setup\cite{Schoelkopf1998a} and driven at resonance ($f_0=367.5$ MHz) by a small ($\leq -95$ dBm) carrier signal. The reflected signal is amplified at 4 K and room temperature, and its amplitude and phase recorded.

We first present measurements on the LR device, showing amplitude (fig.\ 1c) and phase (fig.\ 1d) of the reflected RF signal as a function of gate potentials $V_L$ and $V_R$. In fig.\ 1d we highlight the honeycomb stability pattern characteristic of coupled double dots\cite{VanDerWiel2002}. Each cell corresponds to a particular charge configuration of the double dot, which we label with an offset charge from an arbitrarily chosen (m,n) state. Close to the degeneracy between (m+1,n) and (m,n+1) charge states, it is possible for an electron to tunnel between islands in response to the r.f.\ drive.  This leads to a Sisyphus impedance, and a change in both phase and amplitude is observed. At other charge state boundaries (for example between (m,n) and (m,n+1)), a much smaller Sisyphus impedance is seen because tunnelling is only weakly correlated with the r.f.\ drive due to the much larger resistance of the lead tunnel junctions.

\begin{figure}
\includegraphics{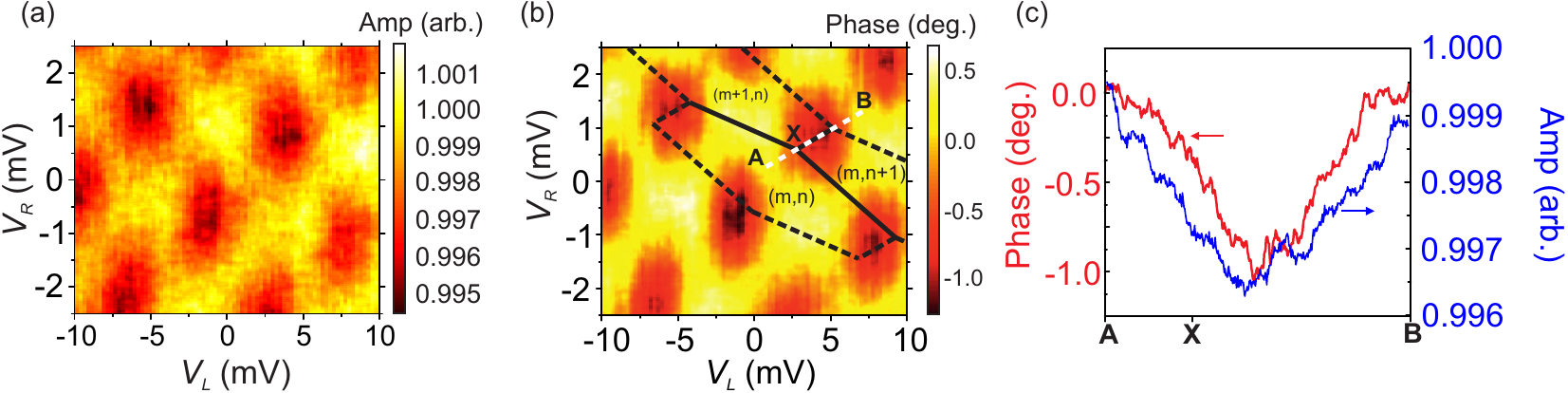}
\caption{(a) The amplitude of the reflected signal for the HR device as a function of $V_L$ and $V_R$. (b) The phase of the reflected signal as a function of $V_L$ and $V_R$. The boundaries of three cells of the charge stability diagram are highlighted. (c) Amplitude and phase along the crossection \textsf{\textbf{A}} $\rightarrow$ \textsf{\textbf{B}}, averaged over 100 measurements.}
\label{fig2}
\end{figure}

We now describe measurements on the HR device. In fig.\ 2a we show phase as a function of $V_L$ and $V_R$, and highlight three charge state cells. We note that the sign of the phase shift induced by $Z_{S}^{Im}$ is dependent upon the junction resistance and the r.f. drive frequency\cite{Ciccarelli2011}, and here is opposite to that for the LR sample. From the stability diagram and bias dependency measurements, we estimate the island charging energies to be $E_{c1} \approx E_{c2} \sim 230$ $\mu$eV, and the electrostatic coupling energy to be $E_{cm} \sim 140$ $\mu$eV.

To probe the tunnelling dynamics of the device, we fix the gate voltages at a point close to the triple point between the (m,n), (m+1,n) and (m,n+1) charge regions, and concentrate on the phase response.  We then take a long (50 s) time trace. In fig.\ 3a a typical trace segment is shown, for $V_{ds} = 0$ mV and $T = 35$ mK. We see a stochastic switching between two phases, separated by $1.4^{\circ}$. From the $S_{11}$ of the resonant circuit, we deduce a corresponding change in $1/f_0\,Z_{S}^{Im}$ of $33$ aF between the two states.

We attribute this impedance change to the thermally driven tunnelling of a single charge through the highly resistive tunnel junctions of the leads. When the device is in the (m,n) state (or other state with even parity relative to this state), the r.f.\ stimulus is unable to drive an electron through the middle junction, and so no Sisyphus impedance is observed (fig.\ 3b, left panel). The addition of an electron to the double island, however, places the device in either the (m+1,n) or (m,n+1) state (or another odd relative parity state), and the Sisyphus impedance is now present due to the extra electron (fig.\ 3b, right panel).

\begin{figure}
\includegraphics{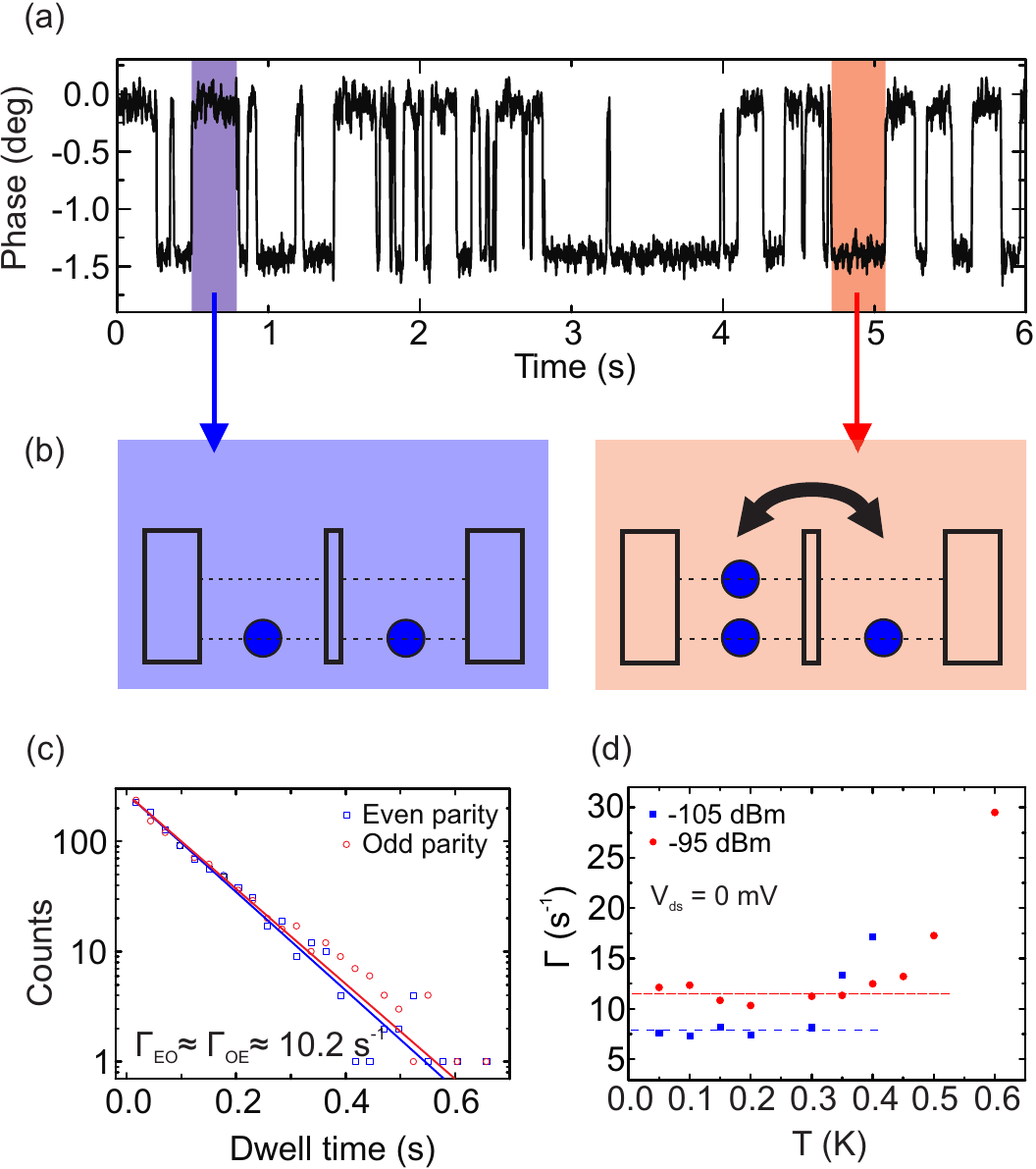}
\caption{(a) Unaveraged phase close to the triple point (X) as a function of time. Stochastic switching is observed between states with an impedance contribution from $Z_{S}^{Im}$ (blue region, left panel in b) and states where this contribution is not present (red region, right panel in b). (b) The Sisyphus process is blocked at even relative parity (left panel), where no excess electron is available, and present at odd relative parity (right panel), where an electron can tunnel between the two islands. (c) Histogram of dwell times for a long time trace at a triple point. We fit exponential distributions (solid lines) and determine both transition rates. (d) Dependence of rate on mixing chamber temperature for carrier powers of $P_{rf} = -105$ dBm (squares) and $P_{rf} = -95$ dBm (circles). The cycle rate saturates (dotted lines) at temperatures below the power dependent electron temperature.}
\label{fig3}
\end{figure}

To determine the tunnel rates of electrons on to and off the device, a time trace is divided into `high' (where there is no Sisyphus impedance) and `low' (where the Sisyphus impedance is present) capacitance periods. The dwell times of these are determined, and they are then histogrammed (fig.\ 3c). We fit separate exponential decays, and extract transition rates between the even and odd parity states, $\Gamma_{EO}$ and $\Gamma_{OE}$. In general, the analysis of Poissonian transition rates requires the finite bandwidth of the measurement setup (in this case $\sim15$ kHz) to be considered\cite{Naaman2006}, but here we note that the measured rates are more than two orders of magnitudes lower than the bandwidth, and so this correction is negligible.

The average rate of parity change is given by $\Gamma = \frac{1}{2}(\Gamma_{EO} + \Gamma_{OE})$. In fig.\ 3d we show $\Gamma$ as a function of the temperature of the dilution fridge mixing chamber for two different r.f.\ carrier powers, with $V_{ds} = 0$. We observe a constant cycle rate for temperatures up to 250 mK ($P_{rf} = -105$ dBm) and 325 mK ($P_{rf} = -95$ dBm). At these temperatures the electron temperature begins to increase and $\Gamma$, which is thermally driven, begins to rise. At lower r.f.\ powers and higher temperatures the signal to noise ratio (SNR) degrades such that the two phase levels cannot be reliably distinguished.

\begin{figure}
\includegraphics{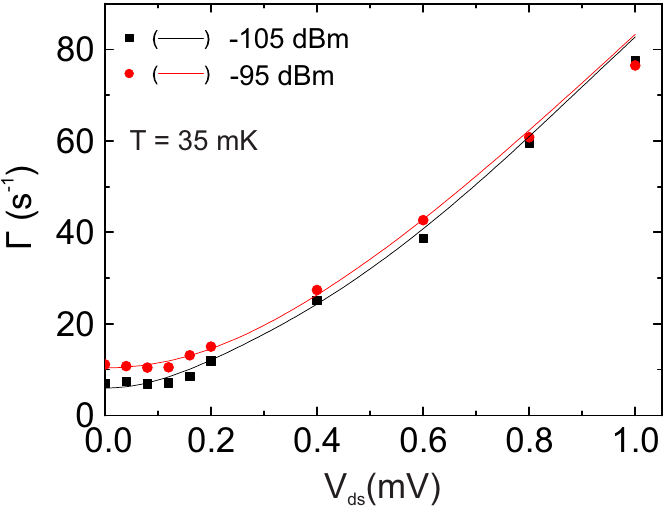}
\caption{Dependence of cycle rate on source-drain bias for carrier powers of $P_{rf} = -105$ dBm (squares) and $P_{rf} = -95$ dBm (circles). The solid lines are fits to the model described in the text.}
\label{fig4}
\end{figure}

We now quantify the behaviour of the charge parity ammeter by measuring $\Gamma$ as a function of $V_{ds}$. We show data at $T = 35$ mK and for two carrier powers (Fig 4). Two regimes can be identified. At low bias ($V_{ds} < 0.1$ mV) $\Gamma$ is dominated by tunnelling events driven both thermally and by the r.f.\ carrier signal. The net charge flow in this regime is significantly less than $\frac{1}{2}e$. At higher bias ($V_{ds} > 0.2$ mV) events in which electrons are transferred from source to the device, or the device to the drain, dominate. This regime extends to lower bias for reduced carrier power.

We model this behaviour by considering individual tunnel events onto and off the island pair. The tunnel rate through a single NIN junction is given by\cite{Grabert1992}

\begin{align*}
\Gamma_{t}=\frac{R_{k}}{R_{T}}\frac{\Delta E/h}{1-\exp(\Delta E/k_{B}T_{E})}
\end{align*}

where $\Delta E$ is the change in chemical potential, $R_k = \frac{h}{e^2}$ and $T_e$ is the electron temperature. There are two contributions to $\Delta E$ for each junction; a d.c. contribution, which includes $V_{ds}$ and the chemical potential of the charge state involved, and an a.c. contribution from the r.f.\ drive.  We include the r.f.\ drive by integrating the tunnel rate over one drive period and renormalising. For forward transport

\begin{align}
\Gamma_{a}=\frac{\omega}{2 \pi}\frac{R_k}{R_t}\int^{\frac{2\pi}{\omega}}_{0}\frac{\Delta E(t)/h}{1-\exp(\Delta E(t)/k_{B}T_{E})}\mathrm{d}t
\end{align}

with $\Delta E(t)=(V_{r.f.} \sin(\omega t)+V_{d.c.})e$, and $V_{r.f.}$ determined from the carrier power and Q factor of the resonant circuit.

To determine the rate of parity change for a given $V_{ds}$ we solve the master equation for the double dot charge states to determine populations in each charge state. The total charge on the island can be increased (decreased) by the tunnelling of an electron from (to) either the source or drain. The rates for these processes are determined by solving eqn 1 numerically for each state. The total rate of parity change is then given by the rate for each charge state weighted by the population of that state.

This description remains valid if a large source-drain bias such that ($e V_{ds} > Ec)$ is applied to the device, and it is therefore no longer in the Coulomb blockade regime.  In this regime charge states other than (m,n), (m+1,n) and (m,n+1) can be occupied and the presence of the Sisyphus impedance depends upon the charge state being odd parity relative to (m,n). By measuring the phase shift, we therefore monitor changes in the charge parity of the device in real time, and at large $V_{ds}$ each switch of parity is due to a net charge transfer of $\frac{1}{2}e$ from drain to source.

In fig.\ 4 we show a fit (solid lines) to the measured rates. The electron temperature is determined as above, and we find a good fit for resistances of $54$ T$\Omega$ for each lead-island tunnel junction, for both r.f.\ powers.

We measured $\Gamma_{EO,OE}$ tunnel rates in the range 10-100 Hz, corresponding to electrical currents of order aA. This demonstrates proof-of-principle of the charge parity ammeter, and we now discuss the prospects of measuring larger currents, as required for example in current metrology. In our experiment, the rates were limited by the high lead-dot tunnel resistances and we could simply decrease these resistances to achieve higher currents. However, we cannot do this indefinitely as the signal to noise ratio will start to limit measurable currents: as an example we were not able to measure the discrete switching in the 5 M$\Omega$ LR device. From the data in fig 2c we find an SNR of 7.3 at a measurement bandwidth of 15 kHz. A sensitivity of $5.3\times10^{-3} e/\sqrt{Hz}$ is therefore implied for the present measurement. There are ways to increase bandwidth: by using higher charging energy devices allowing a higher amplitude r.f.\ carrier signal; by optimising the signal from the Sisyphus impedance; by using a lower noise temperature r.f.\ amplifier (at present $T_{N} = 10$ K); and by using a low-loss superconducting resonant circuit. With these modifications it seems feasible to measure $\Gamma$ in the MHz range, corresponding to pA electrical currents.

In conclusion, we have performed r.f.\ reflectometry measurements on a high resistance aluminium double dot. In measuring the electrical impedance of the device itself we avoided the need for external charge detection in single electron ammetery, and could directly determine the relative charge parity of a metal double dot. This configuration benefits from simplicity in the design, and also is well suited to the use of high bandwidth electrical techniques, in principle enabling relatively large currents to be measured. A related charge parity measurement has application in measuring the spin-state of semiconductor quantum dots\cite{Schroer2012}.

A.J.F. acknowledges support from the Hitachi Research Fellowship and Hitachi Cambridge Laboratory. This work was funded by EPSRC grant EP/H016872/1.

\bibliography{SEA}

\providecommand*\mcitethebibliography{\thebibliography}
\csname @ifundefined\endcsname{endmcitethebibliography}
  {\let\endmcitethebibliography\endthebibliography}{}
\begin{mcitethebibliography}{28}
\providecommand*\natexlab[1]{#1}
\providecommand*\mciteSetBstSublistMode[1]{}
\providecommand*\mciteSetBstMaxWidthForm[2]{}
\providecommand*\mciteBstWouldAddEndPuncttrue
  {\def\EndOfBibitem{\unskip.}}
\providecommand*\mciteBstWouldAddEndPunctfalse
  {\let\EndOfBibitem\relax}
\providecommand*\mciteSetBstMidEndSepPunct[3]{}
\providecommand*\mciteSetBstSublistLabelBeginEnd[3]{}
\providecommand*\EndOfBibitem{}
\mciteSetBstSublistMode{f}
\mciteSetBstMaxWidthForm{subitem}{(\alph{mcitesubitemcount})}
\mciteSetBstSublistLabelBeginEnd
  {\mcitemaxwidthsubitemform\space}
  {\relax}
  {\relax}

\bibitem[Bylander et~al.(2004)Bylander, Duty, and Delsing]{Bylander2004}
Bylander,~J.; Duty,~T.; Delsing,~P. \emph{Nature} \textbf{2004}, \emph{434},
  9\relax
\mciteBstWouldAddEndPuncttrue
\mciteSetBstMidEndSepPunct{\mcitedefaultmidpunct}
{\mcitedefaultendpunct}{\mcitedefaultseppunct}\relax
\EndOfBibitem
\bibitem[Lu et~al.(2003)Lu, Ji, Pfeiffer, West, and Rimberg]{Lu2003}
Lu,~W.; Ji,~Z.; Pfeiffer,~L.; West,~K.~W.; Rimberg,~A.~J. \emph{Nature}
  \textbf{2003}, \emph{423}, 422--5\relax
\mciteBstWouldAddEndPuncttrue
\mciteSetBstMidEndSepPunct{\mcitedefaultmidpunct}
{\mcitedefaultendpunct}{\mcitedefaultseppunct}\relax
\EndOfBibitem
\bibitem[G\"{u}ttinger et~al.(2011)G\"{u}ttinger, Seif, Stampfer, Capelli,
  Ensslin, and Ihn]{Guttinger2011}
G\"{u}ttinger,~J.; Seif,~J.; Stampfer,~C.; Capelli,~A.; Ensslin,~K.; Ihn,~T.
  \emph{Phys. Rev. B} \textbf{2011}, \emph{83}\relax
\mciteBstWouldAddEndPuncttrue
\mciteSetBstMidEndSepPunct{\mcitedefaultmidpunct}
{\mcitedefaultendpunct}{\mcitedefaultseppunct}\relax
\EndOfBibitem
\bibitem[Keller(2008)]{Keller2008}
Keller,~M.~W. \emph{Metrologia} \textbf{2008}, \emph{45}, 102--109\relax
\mciteBstWouldAddEndPuncttrue
\mciteSetBstMidEndSepPunct{\mcitedefaultmidpunct}
{\mcitedefaultendpunct}{\mcitedefaultseppunct}\relax
\EndOfBibitem
\bibitem[Giblin et~al.(2010)Giblin, Wright, Fletcher, Kataoka, Pepper, Janssen,
  Ritchie, Nicoll, Anderson, and Jones]{Giblin2010}
Giblin,~S.~P.; Wright,~S.~J.; Fletcher,~J.~D.; Kataoka,~M.; Pepper,~M.;
  Janssen,~T. J. B.~M.; Ritchie,~D.~A.; Nicoll,~C.~A.; Anderson,~D.; Jones,~G.
  A.~C. \emph{New J. Phys.} \textbf{2010}, \emph{12}, 073013\relax
\mciteBstWouldAddEndPuncttrue
\mciteSetBstMidEndSepPunct{\mcitedefaultmidpunct}
{\mcitedefaultendpunct}{\mcitedefaultseppunct}\relax
\EndOfBibitem
\bibitem[Keller et~al.(1996)Keller, Martinis, Zimmerman, and
  Steinbach]{Keller1996}
Keller,~M.~W.; Martinis,~J.~M.; Zimmerman,~N.~M.; Steinbach,~A.~H. \emph{Appl.
  Phys. Lett.} \textbf{1996}, \emph{69}, 1804\relax
\mciteBstWouldAddEndPuncttrue
\mciteSetBstMidEndSepPunct{\mcitedefaultmidpunct}
{\mcitedefaultendpunct}{\mcitedefaultseppunct}\relax
\EndOfBibitem
\bibitem[Fricke et~al.(2011)Fricke, Hohls, Ubbelohde, Kaestner, Kashcheyevs,
  Leicht, Mirovsky, Pierz, Schumacher, and Haug]{Fricke2011}
Fricke,~L.; Hohls,~F.; Ubbelohde,~N.; Kaestner,~B.; Kashcheyevs,~V.;
  Leicht,~C.; Mirovsky,~P.; Pierz,~K.; Schumacher,~H.~W.; Haug,~R.~J.
  \emph{Phys. Rev. B} \textbf{2011}, \emph{83}, 193306\relax
\mciteBstWouldAddEndPuncttrue
\mciteSetBstMidEndSepPunct{\mcitedefaultmidpunct}
{\mcitedefaultendpunct}{\mcitedefaultseppunct}\relax
\EndOfBibitem
\bibitem[Yamahata et~al.(2011)Yamahata, Nishiguchi, and Fujiwara]{Yamahata2011}
Yamahata,~G.; Nishiguchi,~K.; Fujiwara,~A. \emph{Appl. Phys. Lett.}
  \textbf{2011}, \emph{98}, 222104\relax
\mciteBstWouldAddEndPuncttrue
\mciteSetBstMidEndSepPunct{\mcitedefaultmidpunct}
{\mcitedefaultendpunct}{\mcitedefaultseppunct}\relax
\EndOfBibitem
\bibitem[Fujisawa et~al.(2006)Fujisawa, Hayashi, Tomita, and
  Hirayama]{Fujisawa2006}
Fujisawa,~T.; Hayashi,~T.; Tomita,~R.; Hirayama,~Y. \emph{Science}
  \textbf{2006}, \emph{312}, 1634--1636\relax
\mciteBstWouldAddEndPuncttrue
\mciteSetBstMidEndSepPunct{\mcitedefaultmidpunct}
{\mcitedefaultendpunct}{\mcitedefaultseppunct}\relax
\EndOfBibitem
\bibitem[Fujisawa et~al.(2004)Fujisawa, Hayashi, Hirayama, Cheong, and
  Jeong]{Fujisawa2004}
Fujisawa,~T.; Hayashi,~T.; Hirayama,~Y.; Cheong,~H.~D.; Jeong,~Y.~H.
  \emph{Appl. Phys. Lett.} \textbf{2004}, \emph{84}, 2343\relax
\mciteBstWouldAddEndPuncttrue
\mciteSetBstMidEndSepPunct{\mcitedefaultmidpunct}
{\mcitedefaultendpunct}{\mcitedefaultseppunct}\relax
\EndOfBibitem
\bibitem[Schleser et~al.(2004)Schleser, Ruh, Ihn, Ensslin, Driscoll, and
  Gossard]{Schleser2004}
Schleser,~R.; Ruh,~E.; Ihn,~T.; Ensslin,~K.; Driscoll,~D.~C.; Gossard,~A.~C.
  \emph{Appl. Phys. Lett.} \textbf{2004}, \emph{85}, 2005\relax
\mciteBstWouldAddEndPuncttrue
\mciteSetBstMidEndSepPunct{\mcitedefaultmidpunct}
{\mcitedefaultendpunct}{\mcitedefaultseppunct}\relax
\EndOfBibitem
\bibitem[Vandersypen et~al.(2004)Vandersypen, Elzerman, Schouten, {Willems Van
  Beveren}, Hanson, and Kouwenhoven]{Vandersypen2004}
Vandersypen,~L. M.~K.; Elzerman,~J.~M.; Schouten,~R.~N.; {Willems Van
  Beveren},~L.~H.; Hanson,~R.; Kouwenhoven,~L.~P. \emph{Appl. Phys. Lett.}
  \textbf{2004}, \emph{85}, 4394\relax
\mciteBstWouldAddEndPuncttrue
\mciteSetBstMidEndSepPunct{\mcitedefaultmidpunct}
{\mcitedefaultendpunct}{\mcitedefaultseppunct}\relax
\EndOfBibitem
\bibitem[Gustavsson et~al.(2007)Gustavsson, Shorubalko, Leturcq, Sch\"{o}n, and
  Ensslin]{Gustavsson2007}
Gustavsson,~S.; Shorubalko,~I.; Leturcq,~R.; Sch\"{o}n,~S.; Ensslin,~K.
  \emph{Appl. Phys. Lett.} \textbf{2007}, \emph{92}, 152101\relax
\mciteBstWouldAddEndPuncttrue
\mciteSetBstMidEndSepPunct{\mcitedefaultmidpunct}
{\mcitedefaultendpunct}{\mcitedefaultseppunct}\relax
\EndOfBibitem
\bibitem[Schoelkopf(1998)]{Schoelkopf1998a}
Schoelkopf,~R.~J. \emph{Science} \textbf{1998}, \emph{280}, 1238--1242\relax
\mciteBstWouldAddEndPuncttrue
\mciteSetBstMidEndSepPunct{\mcitedefaultmidpunct}
{\mcitedefaultendpunct}{\mcitedefaultseppunct}\relax
\EndOfBibitem
\bibitem[Qin and Williams(2006)Qin, and Williams]{Qin2006}
Qin,~H.; Williams,~D.~A. \emph{Appl. Phys. Lett.} \textbf{2006}, \emph{88},
  203506\relax
\mciteBstWouldAddEndPuncttrue
\mciteSetBstMidEndSepPunct{\mcitedefaultmidpunct}
{\mcitedefaultendpunct}{\mcitedefaultseppunct}\relax
\EndOfBibitem
\bibitem[Reilly et~al.(2007)Reilly, Marcus, Hanson, and Gossard]{Reilly2007}
Reilly,~D.~J.; Marcus,~C.~M.; Hanson,~M.~P.; Gossard,~A.~C. \emph{Appl. Phys.
  Lett.} \textbf{2007}, \emph{91}, 162101\relax
\mciteBstWouldAddEndPuncttrue
\mciteSetBstMidEndSepPunct{\mcitedefaultmidpunct}
{\mcitedefaultendpunct}{\mcitedefaultseppunct}\relax
\EndOfBibitem
\bibitem[Persson et~al.(2010)Persson, Wilson, Sandberg, Johansson, and
  Delsing]{Persson2010}
Persson,~F.; Wilson,~C.~M.; Sandberg,~M.; Johansson,~G.; Delsing,~P. \emph{Nano
  Lett.} \textbf{2010}, \emph{10}, 953--7\relax
\mciteBstWouldAddEndPuncttrue
\mciteSetBstMidEndSepPunct{\mcitedefaultmidpunct}
{\mcitedefaultendpunct}{\mcitedefaultseppunct}\relax
\EndOfBibitem
\bibitem[Ciccarelli and Ferguson(2011)Ciccarelli, and Ferguson]{Ciccarelli2011}
Ciccarelli,~C.; Ferguson,~A.~J. \emph{New J. Phys.} \textbf{2011}, \emph{13},
  093015\relax
\mciteBstWouldAddEndPuncttrue
\mciteSetBstMidEndSepPunct{\mcitedefaultmidpunct}
{\mcitedefaultendpunct}{\mcitedefaultseppunct}\relax
\EndOfBibitem
\bibitem[Ota et~al.(2010)Ota, Hayashi, Muraki, and Fujisawa]{Ota2010}
Ota,~T.; Hayashi,~T.; Muraki,~K.; Fujisawa,~T. \emph{Appl. Phys. Lett.}
  \textbf{2010}, \emph{96}, 032104\relax
\mciteBstWouldAddEndPuncttrue
\mciteSetBstMidEndSepPunct{\mcitedefaultmidpunct}
{\mcitedefaultendpunct}{\mcitedefaultseppunct}\relax
\EndOfBibitem
\bibitem[Ashoori et~al.(1992)Ashoori, Stormer, Weiner, Pfeiffer, Pearton,
  Baldwin, and West]{Ashoori1992a}
Ashoori,~R.~C.; Stormer,~H.~L.; Weiner,~J.~S.; Pfeiffer,~L.~N.; Pearton,~S.~J.;
  Baldwin,~K.~W.; West,~K.~W. \emph{Phys. Rev. Lett.} \textbf{1992}, \emph{68},
  3088--3091\relax
\mciteBstWouldAddEndPuncttrue
\mciteSetBstMidEndSepPunct{\mcitedefaultmidpunct}
{\mcitedefaultendpunct}{\mcitedefaultseppunct}\relax
\EndOfBibitem
\bibitem[Duty et~al.(2005)Duty, Johansson, Bladh, Gunnarsson, Wilson, and
  Delsing]{Duty2005}
Duty,~T.; Johansson,~G.; Bladh,~K.; Gunnarsson,~D.; Wilson,~C.; Delsing,~P.
  \emph{Phys. Rev. Lett.} \textbf{2005}, \emph{95}\relax
\mciteBstWouldAddEndPuncttrue
\mciteSetBstMidEndSepPunct{\mcitedefaultmidpunct}
{\mcitedefaultendpunct}{\mcitedefaultseppunct}\relax
\EndOfBibitem
\bibitem[Petersson et~al.(2010)Petersson, Smith, Anderson, Atkinson, Jones, and
  Ritchie]{Petersson2010}
Petersson,~K.~D.; Smith,~C.~G.; Anderson,~D.; Atkinson,~P.; Jones,~G. A.~C.;
  Ritchie,~D.~A. \emph{Nano Lett.} \textbf{2010}, \emph{10}, 2789--93\relax
\mciteBstWouldAddEndPuncttrue
\mciteSetBstMidEndSepPunct{\mcitedefaultmidpunct}
{\mcitedefaultendpunct}{\mcitedefaultseppunct}\relax
\EndOfBibitem
\bibitem[Schroer et~al.(2012)Schroer, Jung, Petersson, and Petta]{Schroer2012}
Schroer,~M.~D.; Jung,~M.; Petersson,~K.~D.; Petta,~J.~R. \emph{Phys. Rev.
  Lett.} \textbf{2012}, \emph{109}, 166804\relax
\mciteBstWouldAddEndPuncttrue
\mciteSetBstMidEndSepPunct{\mcitedefaultmidpunct}
{\mcitedefaultendpunct}{\mcitedefaultseppunct}\relax
\EndOfBibitem
\bibitem[Fulton and Dolan(1987)Fulton, and Dolan]{Fulton1987}
Fulton,~T.; Dolan,~G. \emph{Phys. Rev. Lett.} \textbf{1987}, \emph{59},
  109--112\relax
\mciteBstWouldAddEndPuncttrue
\mciteSetBstMidEndSepPunct{\mcitedefaultmidpunct}
{\mcitedefaultendpunct}{\mcitedefaultseppunct}\relax
\EndOfBibitem
\bibitem[van~der Wiel et~al.(2002)van~der Wiel, {De Franceschi}, Elzerman,
  Fujisawa, Tarucha, and Kouwenhoven]{VanDerWiel2002}
van~der Wiel,~W.; {De Franceschi},~S.; Elzerman,~J.~M.; Fujisawa,~T.;
  Tarucha,~S.; Kouwenhoven,~L.~P. \emph{Rev. Mod. Phys.} \textbf{2002},
  \emph{75}, 32\relax
\mciteBstWouldAddEndPuncttrue
\mciteSetBstMidEndSepPunct{\mcitedefaultmidpunct}
{\mcitedefaultendpunct}{\mcitedefaultseppunct}\relax
\EndOfBibitem
\bibitem[Naaman and Aumentado(2006)Naaman, and Aumentado]{Naaman2006}
Naaman,~O.; Aumentado,~J. \emph{Phys. Rev. B} \textbf{2006}, \emph{73}, 4\relax
\mciteBstWouldAddEndPuncttrue
\mciteSetBstMidEndSepPunct{\mcitedefaultmidpunct}
{\mcitedefaultendpunct}{\mcitedefaultseppunct}\relax
\EndOfBibitem
\bibitem[Grabert and Devoret(1992)Grabert, and Devoret]{Grabert1992}
Grabert,~H.; Devoret,~M.~H. \emph{{Single charge tunneling: Coulomb blockade
  phenomena in nanostructures}}; Plenum Press: New York, 1992; p~41\relax
\mciteBstWouldAddEndPuncttrue
\mciteSetBstMidEndSepPunct{\mcitedefaultmidpunct}
{\mcitedefaultendpunct}{\mcitedefaultseppunct}\relax
\EndOfBibitem
\end{mcitethebibliography}

\end{document}